\newif\ifabstract
\abstracttrue
 \abstractfalse 
\newif\iffull
\ifabstract \fullfalse \else \fulltrue \fi

\documentclass[11pt]{article}
\usepackage{amsfonts}
\usepackage{amssymb}
\usepackage{amstext}
\usepackage{amsmath}
\usepackage{xspace}
\usepackage{theorem}
\usepackage{graphicx}
\usepackage{url}
\usepackage{graphics}
\usepackage{colordvi}
\usepackage{colordvi}
\usepackage{subfigure}

\advance \oddsidemargin by -0.8in
\newcommand{\myparskip}{3pt}
\parskip \myparskip

        {\hspace*{\fill}$\Box$\par\vspace{4mm}}

\newcommand{\be}{\begin{enumerate}}
\newcommand{\ee}{\end{enumerate}}
\newcommand{\bd}{\begin{description}}
\newcommand{\ed}{\end{description}}
\newcommand{\bi}{\begin{itemize}}
\newcommand{\ei}{\end{itemize}}

\newtheorem{remark}{Remark}[section]
\newenvironment{proof}{\par \smallskip{\bf Proof:}}{\hfill\stopproof}
\def\stopproof{\square}
\def\square{\vbox{\hrule height.2pt\hbox{\vrule width.2pt height5pt \kern5pt
\vrule width.2pt} \hrule height.2pt}}

\renewcommand{\phi}{\varphi}

\mathchardef\hyphen="2D

\begin{document}

\title{Consensus-based joint target tracking and sensor localization}
\author{
	Lin Gao\thanks{School of Electronic Engineering, Center for Cyber Security, University of Electronic Science and Technology of China. Email: {\tt gaolin\_uestc@126.com}. } \and 
	Giorgio Battistelli\thanks{Dipartimento di Ingegneria dell'Informazione (DINFO), Universit\`{a} degli Studi di Firenze. Email: {\tt giorgio.battistelli@unifi.it}. } \and 
	Luigi Chisci\thanks{Dipartimento di Ingegneria dell'Informazione (DINFO), Universit\`{a} degli Studi di Firenze. Email: {\tt luigi.chisci\@unifi.it}.} \and
	Ping Wei\thanks{School of Electronic Engineering, Center for Cyber Security, University of Electronic Science and Technology of China. Email: {\tt pwei@uestc.edu.cn}. }
}

\begin{titlepage}
\maketitle

\thispagestyle{empty}

\begin{abstract}
In this paper, consensus-based Kalman filtering is extended to deal with the problem of joint target tracking and sensor self-localization in a distributed wireless sensor network.
The average weighted Kullback-Leibler divergence, which is a function of the unknown drift parameters, is employed as the cost to measure the discrepancy between the fused posterior distribution and the local distribution at each sensor.
Further, a reasonable approximation of the cost is proposed and an online technique is introduced to minimize the approximated cost function with respect to the drift parameters stored in each node.
The remarkable features of the proposed algorithm are that it needs no additional data exchanges, slightly increased memory space and computational load comparable to the standard consensus-based Kalman filter.
Finally, the effectiveness of the proposed algorithm is demonstrated through simulation experiments on both a tree network and a network with cycles as well as for both linear and nonlinear sensors.
\end{abstract}

\end{titlepage}

\label{--------------------------------------------sec: intro---------------------------------------------------}
\section{Introduction}\label{sec: intro}
Distributed target tracking (DTT) has recently attracted great attention due to the fast development of wireless sensor networks (WSNs).
The remarkable feature of a WSN is that there is no fusion center, and the nearby nodes can freely exchange data.
The goal of DTT is to achieve scalability and comparable performance with respect  to the centralised algorithms\cite{olfati2007consensus}.
In the past, most work on DTT assumed that all the sensors have been correctly localized with respect to a global coordinate system, or at least that each sensor knows the positions of neighbours with respect to its own local coordinate system\cite{kamal2013information,das2015distributed,battistelli2014parallel,cattivelli2010diffusion,cattivelli2010distributed}.
In many practical scenarios, however, sensor self-localization is also an important task to be jointly faced along with target tracking.
Most of the existing work on sensor self-localization relied on two approaches.
In the first approach, called \textit{cooperative localization}, each sensor is provided with direct measurements relative to the positions of its neighbors \cite{vcapkun2002gps,mao2007wireless,frampton2006acoustic,meyer2012simultaneous,sun2012accurate,ihler2005nonparametric,moses2003self,patwari2005locating,shao2014efficient}.
The second approach is based on exploiting some reference nodes of known positions (also called anchors)
in the global coordinate system \cite{khan2009distributed,vemula2009sensor,cevher2006acoustic,chen2011sequential,pedersen2011variational}. The locations of anchors are assumed known a priori or can be obtained by using global localization technology such as, e.g.,
GPS (Global Positioning System).
Undoubtedly both approaches have their limitations.
The former requires additional devices for measuring the positions of the neighboring nodes, and also that such devices have the capability of distinguishing signals
from different nodes, or otherwise a data association problem has to be solved.
The latter can only be used in some specific scenarios where either prior knowledge of the surveillance area is available or signals from the global localization equipment can be received.
In this paper, the interest is for a technique that neither needs sensing the positions of neighbors nor the presence of reference nodes.

In this respect, two interesting techniques for
\textit{joint target tracking and sensor self-localization} (JTTSL) have been presented in  \cite{kantas2012distributed} and  \cite{uney2016cooperative}.
In particular, \cite{kantas2012distributed} exploits online distributed \textit{maximum likelihood} (ML) and \textit{expectation-maximization} (EM) methods.
The nodes iteratively exchange the local likelihoods  based on the message passing (belief propagation) technique.
This approach, however, suffers from three major drawbacks.
First, each node must store the data of all its neighbors and thus needs much extra memory space.
Secondly, at each sampling interval several iterations must be carried out in order to exchange the data through the network.
The third and most important drawback is that the employed message passing method is well suited for networks with tree topology but suffers from the problem of double counting in networks with cycles.

Conversely, \cite{uney2016cooperative} follows a Bayesian approach in order to
compute in each node the posterior distribution of the drift parameters (i.e., the relative positions of the neighbors).
Specifically, a Monte Carlo method is  adopted to represent such a distribution.
The disadvantage of \cite{uney2016cooperative} is, therefore, that it needs a large amount of particles to approximate the drift parameter distribution,
thus implying a heavy computational load which may be unsuitable for sensor nodes with limited computing capabilities.

In this paper, we propose to solve the JTTSL problem exploiting consensus \cite{olfati2004consensus,xiao2005scheme,carli2008distributed,olfati2009kalman,stankovic2009consensus,battistelli2015consensus}. An advantage of a consensus approach is that any individual node needs only to store the integrated information resulting from the fusion with all its neighbors, and thus does not require significant extra memory space.
Furthermore, consensus-based algorithms avoid double counting of information for any kind of network, also involving cycles.
In the proposed algorithm, the weighted Kullback-Leibler (WKL) divergence \cite{battistelli2014kullback} is employed as optimal cost function to measure the
discrepancy between estimated and true drift parameters.
An online optimization method is then used to update the drift parameters at each sampling time.
The advantages of the proposed algorithm are that it needs no additional data exchanges, slight additional memory space and computational load compared to the standard consensus Kalman filter (CSKF), which means that the proposed algorithm can be easily implemented on low-cost sensor nodes and, thus, have
wide potential application.
The proposed algorithm, named \textit{Joint Target Tracking and Sensor Localization CSKF} (JTTSL-CSKF), can also be regarded as a generalized version of the CSKF.

The rest of the paper is organized as follows.
Section II deals with the JTTSL problem formulation.
Section III reviews the CSKF approach to target tracking whenever the locations of neighboring sensor nodes are known.
Then, section IV presents a novel consensus-based approach to JTTSL.
Section V provides simulation examples to demonstrate the effectiveness of the proposed approach.
Section VI ends the paper with concluding remarks as well as perspectives for future work.

\section{Problem Formulation}
This paper addresses the problem of target tracking by means of a distributed sensor network wherein each sensor node gets measurements of the target relative to its local coordinate system.
The sensor network is denoted as $\left( {{\cal N},{\cal A}} \right)$, where ${\cal N}$ is the set of sensor nodes and ${\cal A} \subseteq {\cal N} \times {\cal N}$ the set of connections such that $\left( {i,j} \right) \in {\cal A}$ if node $j$ can receive data from node $i$.
Let $\left| {\cal N} \right|$ denotes the cardinality of ${\cal N}$.
Further, for each node $i \subseteq {\cal N}$, ${{\cal N}^i}$ will denote the set of its in-neighbor nodes (including itself).
It is further assumed that all sensors are time-synchronized.

The state vector of the target, in the local coordinates of node $i$, is denoted as ${\mathop{\rm x}\nolimits} _t^i = [\xi_t^i,\dot{\xi}_t^i,\eta_t^i,\dot{\eta}_t^i]^T$ where $(\xi_t^i,\eta_t^i)$ and
$(\dot{\xi}_t^i,\dot{\eta}_t^i)$ denote the Cartesian coordinates of position and velocity at time $t$, respectively.
The motion of the target is modelled by
\begin{equation}   \label{ds}
{\mathop{\rm x}\nolimits} _{t + 1}^i = {f_t}\left( {{\mathop{\rm x}\nolimits} _t^i} \right) + {w_t}.
\end{equation}

Further, at each sensor $i$, measurements of the target are obtained according to

\begin{equation}   \label{mf}
{\rm y}_t^i = h_t^i\left( {{\mathop{\rm x}\nolimits} _t^i} \right) + v_t^i,i \in {\cal N}.
\end{equation}

It is assumed that ${w_t},v_t^1,v_t^2, \ldots $ are mutually uncorrelated zero-mean white noises with covariances $Q_t = E[{w_t}w_t^T] > 0$ and $R_t^i = E[v_t^i{(v_t^i)^T}] > 0$.
It is also convenient to define the noise information matrices $W_t \buildrel \Delta \over = Q_t^{ - 1}$ and $V_t^i \buildrel \Delta \over = (R_t^i)^{ - 1}$.
Without loss of generality, it is supposed that each node is located at the origin of its local coordinate system.
Let $\left( \xi^{i,j},\eta^{i,j} \right)$ denote the position of node $j$ in the local coordinates of node $i$ and ${\theta ^{i,j}} \buildrel \Delta \over = [{\xi^{i,j}},0,{\eta^{i,j}},0]'$
the drift parameter of sensor $j$ with respect to $i$.
Then, assuming that the sensor nodes are motionless, the target states ${\mathop{\rm x}\nolimits} _t^i$ and ${\mathop{\rm x}\nolimits} _t^j$ in nodes $i$ and $j$ are related by
\begin{equation}   \label{pr}
{\mathop{\rm x}\nolimits} _t^i = {\mathop{\rm x}\nolimits} _t^j + {\theta ^{i,j}}
\end{equation}
Notice that ${\theta ^{i,j}}{\rm{ =  - }}{\theta ^{j,i}}$ and  ${\theta ^{i,i}} = 0$.
Due to the fact that each sensor can only communicate with its neighbours, for self-localization purposes only $\theta^{i,j}$  for
$j \in \mathcal{N}^i \backslash \{ i \}$ are needed by node $i$.
Let $\Theta$ denote the vector of all drift parameters, i.e. $\Theta = col \left( {\theta ^{i,j}},\left( {i,j} \right) \in {\cal A} \right)$.
Hence, sensor $i$ is only interested to the sub-vector $\Theta^i = col \left( \theta^{i,j}, j \in \mathcal{N}^i  \backslash \{ i \} \right)$.
The objective of distributed JTTSL is therefore, for each sensor node $i$, to jointly estimate the target state ${\rm x}_t^i$ (in the local coordinates of node $i$) and the drift parameter
vector $\Theta^i$.

\section{Distributed state estimation based on consensus}

Suppose preliminarily that  all drift parameters ${\theta ^{i,j}},\left( {i,j} \right) \in {\cal A}$, are a priori known. Such an assumption will be relaxed in the following section.
Suppose further that in each node $i$ of the network a probability density function (PDF) $p^i$ is available representing the information at node $i$
on the state vector $\rm x$ expressed in local coordinates. If no information on the correlation between the PDFs of different sensors is available,
when fusing the information of different sensors care must be taken in order to avoid the data incest phenomenon, that is the double counting of common
information. 
As well known, fusion rules that are robust to double counting are Covariance Intersection (CI) and Exponential Mixture Density
(the generalization of CI for arbitrary densities).
Specifically, if the PDFs of all nodes are Gaussian with mean $\rm x^i$ and covariance $P^i$, then the fused PDF in the local coordinates of node $i$ is also Gaussian with mean ${\overline {\rm x}^i}$ and covariance $\overline{P}^i$ given by
\begin{equation}   \label{OMC}
\begin{array}{c}
{\left( \overline{P}^i \right)^{ - 1}} = \sum\limits_{j \in {\cal N}} {{\pi ^{i,j}}{{\left( P^j \right)}^{ - 1}}} \\
{\left( \overline{P}^i \right)^{ - 1}}{{\overline {\rm x}}^i} = \sum\limits_{j \in {\cal N}} {{\pi ^{i,j}}{{\left( P^j \right)}^{ - 1}}{({\rm x}^j}+ \theta^{i,j})}
\end{array}
\end{equation}
where, for any $i$, the weights $\pi^{i,j}, j \in \mathcal N$, are such that $\sum_{j \in \mathcal N} \pi^{i,j} = 1$ and $\pi^{i,j} > 0$.
From an information theoretic point of view, such a fusion rule admits a meaningful interpretation in terms of average with respect to the
Kullback-Leibler (KL) divergence  \cite{battistelli2014kullback}. In fact, the PDF
${{\overline p}^i}( \cdot )$ is the one minimizing
the weighted Kullback-Leibler (WKL) divergence in that \cite{battistelli2014kullback}
\begin{equation}   \label{WKL}
{{\overline p}^i}  = \arg \min_{p} \sum\limits_{j \in {\cal N}} {{\pi ^{i,j}}{D_{KL}}\left( {\left. {{{p}}} \right\|{p^{i,j}}} \right)}
\end{equation}
where $p^{i,j}$ denotes the PDF representing the information available at node $j$ expressed in the local coordinates of node $i$ and
${D_{KL}}(\left.  p  \right\| q )$ denotes the KL divergence of $q$ from $p$ defined as
\begin{equation}   \label{KL}
{D_{KL}}\left( \left. p  \right \| q \right) = \int {p({\rm x})\log \frac{{{{p}}({\rm x})}}{{{q}({\rm x})}}d{\rm x}} \, .
\end{equation}

While (\ref{OMC}) and (\ref{WKL}) require access to the PDFs of all the sensors, it was shown in \cite{battistelli2014kullback}
that actually the average PDF can be computed in a distributed way (i.e. exchanging only information with the neighbors) by using consensus.
Combining this idea with local Kalman filters running in each network node, it is possible to derive a consensus Kalman filter with guaranteed
stability properties.

Specifically, let us denote the local target estimated state vector and covariance at node $i$ and time $t$ as $\hat{\rm{x}}_{t|t}^i$ and
$P_{\left. t \right|t}^i$. The local information matrices are denoted as

\begin{equation}   \label{IF}
\Omega _{\left. t \right|t - 1}^i \buildrel \Delta \over = {(P_{\left. t \right|t - 1}^i)^{ - 1}},\;\;\;\Omega _{\left. t \right|t}^i \buildrel \Delta \over = {(P_{\left. t \right|t}^i)^{ - 1}}
\end{equation}
and the local information vectors

\begin{equation}   \label{IV}
q_{\left. t \right|t - 1}^i \buildrel \Delta \over = \Omega _{\left. t \right|t - 1}^i\hat {\rm x}_{\left. t \right|t - 1}^i,\;\;\;q_{\left. t \right|t}^i \buildrel \Delta \over = \Omega _{\left. t \right|t}^i\hat {\rm x}_{\left. t \right|t}^i
\end{equation}

In the context of distributed target tracking, each node uses the received measurements to first update the local information, and then exchanges data with the neighbors so that information fusion is carried out in order to improve the tracking performance of each node.
In the information fusion step, the consensus algorithm can be used to minimize
(\ref{WKL}).
The CSKF is summarized in Table \ref{KFCI}.
Note that the consensus weights ${\pi ^{i,j}}$ should satisfy ${\pi ^{i,j}} > 0$ and $\sum\nolimits_{j \in {{\cal N}^i}} {{\pi ^{i,j}}}  = 1$.
For the choice of the consensus weights ${\pi ^{i,j}}$, which is outside the scope of this paper, please refer to \cite{battistelli2015consensus}.

In terms of KL average, each consensus step admits the following interpretation.
Let $p^{i,j}_t(\rm x,\ell)$ denote the PDF available at node $j$ at consensus step $\ell$ of the $t$-th sample interval, expressed in the local coordinates
of node $i$. Then, assuming Gaussian PDFs\footnote{Clearly, the PDFs are Gaussian only when both the motion and measurement models
	are linear and both process and measurement noises are Gaussian. Otherwise this amounts to approximating the true local PDF with a Gaussian one having same mean and covariance.}, the new PDF $p^{i,i}_t(\rm x,\ell+1)$ at node $i$ is the one minimizing the weighted KL divergence with respect to the neighbors in that
\begin{equation}   \label{WKLi}
p^{i,i}_t({\rm x},\ell+1)  = \arg \min_{p} \sum\limits_{j \in {{\cal N}^i}} {{\pi^{i,j}{D_{KL}}\left( {\left. {{p}} \right\|{p^{i,j}_t(\cdot,\ell)}} \right)}} \, .
\end{equation}

\begin{table}
	\caption{Consensus Kalman Filter (node $i$, time $t$)}   \label{KFCI}
	\begin{center}
		\begin{tabular}{l p{5cm}|}
			\hline \hline
			\underline{Correction} \\ \hline
			$\;\;\;\;$  $C_t^i = {\left. {\frac{{\partial {h_t^i}}}{{\partial {{\rm x}_t}}}} \right|_{{{\rm x}_t} = \hat {\rm x}_{t|t-1}^i}}$    \\
			$\;\;\;\;$  $\Omega _{t|t}^i(0) = \Omega _{t|t - 1}^i + (C_t^i)^T V_t^i C_t^i$ \\
			$\;\;\;\;$  Sample the local measurement ${\rm y}_t^i$, then \\
			$\;\;\;\;$  $\bar {\rm y}_t^i = {\rm y}_t^i - {h_t}\left( {\hat {\rm x}_{t|t-1}^i} \right) + C_t^i \hat {\rm x}_{t|t-1}^i$    \\
			$\;\;\;\;$  $q_{t|t}^i(0) = q_{t|t - 1}^i + (C_t^i)^T V_t^i\bar {\rm y}_t^i$  \\
			\hline
			\underline{Consensus} \\ \hline
			$\;\;$ For $\ell = 0,1, \ldots ,L - 1$ do \\
			$\;\;\;\;$  Exchange the information with its neighbors and \\
			fuse the quantities $q_t^i(\ell)$ and $\Omega _t^i(\ell)$ according to \\
			$\;\;\;\;\;\;$ $q_t^i(\ell + 1) = \sum\limits_{j \in {{\cal N}^i}} {{\pi ^{i,j}}[q_{\left. t \right|t}^j(\ell) + \Omega _{\left. t \right|t}^j(\ell){\theta ^{i,j}}]}$  \\
			$\;\;\;\;\;\;$ $\Omega _t^i(\ell + 1) = \sum\limits_{j \in {{\cal N}^i}} {{\pi ^{i,j}}\Omega _{\left. t \right|t}^j(\ell)}$  \\
			$\;\;$ End for \\
			$\;\;$ $\hat {\rm x}_{\left. t \right|t}^i = {[\Omega _{\left. t \right|t}^i(L)]^{ - 1}}q_t^i(L),\;\;\Omega _{\left. t \right|t}^i = \Omega _t^i(L)$  \\ \hline
			\underline{Prediction}  \\ \hline
			$\;\;$ $\hat {\rm x}_{\left. {t + 1} \right|t}^i = {f_t}\left( {\hat {\rm x}_{\left. t \right|t}^i} \right),\; A_t = {\left. {\frac{{\partial {f_t}}}{{\partial {{\rm x}_t}}}} \right|_{{{\rm x}_t} = \hat {\rm x}_{\left. t \right|t}^i}}$  \\
			$\;\;$ $\Omega_{t+1|t}^i = W_t - W_t A_t \left( \Omega _{t|t}^i + A_t^T W_t A_t \right)^{ - 1} A_t^T W_t$  \\
			$\;\;$ $q_{\left. {t + 1} \right|t}^i = \Omega _{\left. {t + 1} \right|t}^i \hat {\rm x}_{\left. {t + 1} \right|t}^i$ \\
			\hline \hline
		\end{tabular}
	\end{center}
\end{table}

\section{Joint state estimation and online drift parameters calibration}

In real applications, the drift parameters ${\theta ^{i,j}},\left( {i,j} \right) \in {\cal A}$, are not always a priori known and need, therefore, be estimated together
with the state vector $\rm x_t$.
Notice preliminarily that a possible solution would amount to considering the distributed estimation problem for a global state $X_t$ consisting of the target kinematic state $\rm x_t$ and the overall drift parameter vector $\Theta$ containing the drifts parameters of all links in the network.
However, such an approach would not be scalable since
the dimension of the global state $X_t = \left[ {\rm x}_t^T, \Theta^T \right]^T$ would increase with the number of network links and further would require the knowledge of the network topology.
Hence, in this paper we follow a different approach in which each node only estimates the drift parameters $\Theta^i$ pertaining to its neighbors $j \in \mathcal N^i \backslash \{i \}$.
This is a suboptimal approach which, however, has the advantage of being scalable and does not require any global information on the network topology.
Further, in order to keep the communication load as low as possible, we propose an approach that does not require any additional data exchange
with respect to the Consensus Kalman Filter of Table I.

The idea is to exploit the information theoretic interpretation of the consensus step  (\ref{WKLi}) in order to define a suitable loss function
which can be used for estimation of the drift parameters. In fact, since the transformation from the local coordinates of node $j$ to those of node $i$
is a function of the drift parameter $\theta^{i,j}$, it is possible to define the loss function
\begin{equation}\label{eq:loss}
J^i_t(\Theta^i , \ell) =  \min_{p} \sum\limits_{j \in {{\cal N}^i}} {{\pi^{i,j}{D_{KL}}\left( {\left. {{p}} \right\|{p^{i,j}_t(\cdot,\ell)}} \right)}}
\end{equation}
where the vector $\Theta^i$ contains all drift parameters relative to the neighbors of node $i$.
The rationale for such a choice is that, when all the local filters perform well, then
all the local PDFs should provide a reasonably accurate estimate of the target state in local coordinates. In this case, the discrepancy between the
PDFs in two neighboring nodes $i$ and $j$ is mainly due to the different coordinates.
Hence, it is reasonable to take as estimate of the drift parameter
$\theta^{i,j}$ the value which minimizes such a discrepancy.
In turn, the estimates of the drift parameters are used for the distributed state estimator.

Assuming Gaussian PDFs, the loss function (\ref{eq:loss}) turns out to be a quadratic function of $\Theta^i$. In fact,
for two Gaussian distributions ${p^i(\cdot)},{p^j(\cdot)}$ with means ${{\rm x}^i},{{\rm x}^j} \in \mathbb{R}^d$ and covariance matrices $P^i,P^j$,
their KL divergence in the coordinates of node $i$ can be written as
\begin{equation}   \label{GKL}
\begin{array}{l}
{D_{KL}}(\left. {{p_i}} \right\|{p_j}) = \frac{1}{2}\left[ {{{\left( {{{\rm x}^i} - \left( {{{\rm x}^j} + {\theta ^{i,j}}} \right)} \right)}^T}{{\left( P^j \right)}^{ - 1}}\left( {{{\rm x}^i} - \left( {{{\rm x}^j} + {\theta ^{i,j}}} \right)} \right)} \right.
\left. { - d + tr \left( \left( P^j \right)^{ - 1} P^i \right) + \ln \left( {\frac{\det P^j}{\det P^i}} \right)} \right].
\end{array}
\end{equation}
Then, the PDF minimizing the WKL in (\ref{eq:loss}) is again Gaussian with mean
\begin{equation}\label{x:tilde}
\tilde{\rm x}_t^i(\Theta^i,\ell+1) = [  \Omega_t^i(\ell+1) ]^{-1} \sum_{j \in \mathcal N^i} \pi^{i,j}  \Omega^j_t (\ell)  [\hat {\rm x}_t^j (\ell) + \theta^{i,j}] \, ,
\end{equation}
$\hat {\rm x}_t^j (\ell) = [\Omega^j_t (\ell)]^{-1} q^j_t(\ell)$, and inverse covariance
$\Omega_t^i(\ell+1)$ independent of $\Theta^i$.

In view of (\ref{GKL}) and (\ref{x:tilde}), it is an easy matter to see that the loss function $J^i_t(\Theta^i , \ell) $ can be written as
the quadratic form
\begin{equation}  \label{OCF}
J^i_t(\Theta^i , \ell) =  (\Theta^i)^T \Phi^i_t(\ell)  \Theta^i + 2 (\Theta^i)^T  \varphi^i_t(\ell) + c^i_t(\ell)
\end{equation}
for suitable $\Phi^i_t(\ell), \varphi^i_t(\ell), c^i_t(\ell)$ independent of $\Theta^i$.
In particular, it can be checked that

\begin{align}
\Phi _t^i\left( \ell  \right) &= \frac{1}{2}\left[ {\Psi _t^i\left( \ell  \right){E^i}{{\left( {\Omega _t^i\left( {\ell  + 1} \right)} \right)}^{ - 1}}{{\left( {{E^i}} \right)}^T}\Psi _t^i\left( \ell  \right) - \Psi _t^i\left( \ell  \right)} \right] \label{OmeA}  \\
\varphi _t^i\left( \ell  \right) &= \Phi _t^i\left( \ell  \right) \left[ {{\bf{x}}_t^i\left( \ell  \right) - {E^i}{\rm{\hat x}}_t^i\left( \ell  \right)} \right]  \label{PsiA}
\end{align}
where 

\begin{align}
\Psi _t^i\left( \ell  \right) &= {\mathop{ block-diag}\nolimits} \left( {{\pi ^{i,j}}\Omega _t^j\left( \ell  \right),j \in {{\cal N}^i}\backslash \left\{ i \right\}} \right)  \label{PsiB} \\
E^i &= col\underbrace {\left( {{I_4}, \cdots ,{I_4}} \right)}_{\left| {{{\cal N}^i}} \right| - 1\;times} \label{Matrixone} \\
{\bf{x}}_t^i\left( \ell  \right) &= col\left( {\hat {\rm x}_t^j\left( \ell  \right),j \in {{\cal N}^i}\backslash \left\{ i \right\}} \right) \label{InteState}
\end{align}
%
and $block-diag$ denotes the block diagonal matrix obtained from its arguments. 

\begin{proof}
	According to the definitions of (\ref{PsiB}) to (\ref{InteState}), we have
	
	\begin{align*}
	\Omega _t^i\left( {\ell  + 1} \right) &= {\left( {{E^i}} \right)^T}\Psi _t^i\left( \ell  \right){E^i} + {\pi ^{i,i}}\Omega _t^i\left( \ell  \right)  \\
	\tilde {\rm x}_t^i\left( {{\Theta ^i},\ell  + 1} \right) &= {\left( {\Omega _t^i\left( {\ell  + 1} \right)} \right)^{ - 1}}\left[ {{{\left( {{E^i}} \right)}^T}\Psi _t^i\left( \ell  \right)\left( {{\bf{x}}_t^i\left( \ell  \right) + {\Theta ^i}} \right) + {\pi ^{i,i}}\Omega _t^i\left( \ell  \right){\rm{\hat x}}_t^i\left( \ell  \right)} \right]
	\end{align*}
	
	Then the cost $J_t^i({\Theta ^i},\ell)$ is computed as
	
	\begin{align}
		&J_t^i({\Theta ^i},\ell) \nonumber  \\
		&= \sum\limits_{j \in {{\cal N}^i}} {{\pi ^{i,j}}{D_{KL}}\left( {\left. {\tilde p_t^i\left( { \cdot ,\ell  + 1} \right)} \right\|p_t^{i,j}\left( {\cdot,\ell } \right)} \right)}   \nonumber \\
		& = \frac{1}{2}\sum\limits_{j \in {{\cal N}^i}} {\left[ {{{\left( {\tilde {\rm x}_t^i\left( {{\Theta ^i},\ell  + 1} \right) - \left( {{\rm{\hat x}}_t^j\left( \ell  \right) + {\theta ^{i,j}}} \right)} \right)}^T}{\pi ^{i,j}}\Omega _t^j\left( \ell  \right)\left( {\tilde {\rm x}_t^i\left( {{\Theta ^i},\ell  + 1} \right) - \left( {{\rm{\hat x}}_t^j\left( \ell  \right) + {\theta ^{i,j}}} \right)} \right)} \right]}   \nonumber  \\
		& =  \frac{1}{2}\left\{ { {{\left( {\tilde {\rm x}_t^i\left( {{\Theta ^i},\ell  + 1} \right)} \right)}^T}\Omega _t^i\left( {\ell  + 1} \right)\left( {\tilde {\rm x}_t^i\left( {{\Theta ^i},\ell  + 1} \right)} \right)} \right.   \nonumber  \\
		& \;\;\;\; - \left. {\sum\limits_{j \in {{\cal N}^i}\backslash \left\{ i \right\}} {{{\left( {\left( {{\rm{\hat x}}_t^j\left( \ell  \right) + {\theta ^{i,j}}} \right)} \right)}^T}{\pi ^{i,j}}\Omega _t^j\left( \ell  \right)\left( {{\rm{\hat x}}_t^j\left( \ell  \right) + {\theta ^{i,j}}} \right)}  + {{\left( {{\rm{\hat x}}_t^i\left( \ell  \right)} \right)}^T}{\pi ^{i,i}}\Omega _t^i\left( \ell  \right){\rm{\hat x}}_t^i\left( \ell  \right)} \right\} + C   \nonumber \\
		& =  \frac{1}{2}\left\{ {{{\left[ {{{\left( {{E^i}} \right)}^T}\Psi _t^i\left( \ell  \right)\left( {{\bf{x}}_t^i\left( \ell  \right) + {\Theta ^i}} \right) + {\pi ^{i,i}}\Omega _t^i\left( \ell  \right){\rm{\hat x}}_t^i\left( \ell  \right)} \right]}^T}} \right.{\left( {\Omega _t^i\left( {\ell  + 1} \right)} \right)^{ - 1}}  \nonumber  \\
		& \;\;\;\; \times \left[ {{{\left( {{E^i}} \right)}^T}\Psi _t^i\left( \ell  \right)\left( {{\bf{x}}_t^i\left( \ell  \right) + {\Theta ^i}} \right) + {\pi ^{i,i}}\Omega _t^i\left( \ell  \right){\rm{\hat x}}_t^i\left( \ell  \right)} \right] - {\left[ {\left( {{\bf{x}}_t^i\left( \ell  \right) + {\Theta ^i}} \right)} \right]^T}\Psi _t^i\left[ {\left( {{\bf{x}}_t^i\left( \ell  \right) + {\Theta ^i}} \right)} \right]  \nonumber  \\
		& \;\;\;\; \left. { - {{\left( {{\rm{\hat x}}_t^i\left( \ell  \right)} \right)}^T}{\pi ^{i,i}}\Omega _t^i\left( \ell  \right){\rm{\hat x}}_t^i\left( \ell  \right)} \right\} + C  \nonumber  \\
		& = {\left( {{\Theta ^i} + {\bf{x}}_t^i\left( \ell  \right)} \right)^T}\Phi _t^i\left( \ell  \right)\left( {{\Theta ^i} + {\bf{x}}_t^i\left( \ell  \right)} \right) + 2{\left( {{\Theta ^i} + {\bf{x}}_t^i\left( \ell  \right)} \right)^T}\frac{1}{2}\Psi _t^i\left( \ell  \right){E^i}{\left( {\Omega _t^i\left( {\ell  + 1} \right)} \right)^{ - 1}}{\pi ^{i,i}}\Omega _t^i\left( \ell  \right){\rm{\hat x}}_t^i\left( \ell  \right) + C  \nonumber   \\
		& = {\left[ {{\Theta ^i} + {\bf{x}}_t^i\left( \ell  \right) + {{\frac{1}{2}\left( {\Phi _t^i\left( \ell  \right)} \right)}^{ - 1}}\Psi _t^i\left( \ell  \right){E^i}{{\left( {\Omega _t^i\left( {\ell  + 1} \right)} \right)}^{ - 1}}{\pi ^{i,i}}\Omega _t^i\left( \ell  \right){\rm{\hat x}}_t^i\left( \ell  \right)} \right]^T}\Phi _t^i\left( \ell  \right)   \nonumber  \\
		& \;\;\;\; \times \left[ {{\Theta ^i} + {\bf{x}}_t^i\left( \ell  \right) + {{\frac{1}{2}\left( {\Phi _t^i\left( \ell  \right)} \right)}^{ - 1}}\Psi _t^i\left( \ell  \right){E^i}{{\left( {\Omega _t^i\left( {\ell  + 1} \right)} \right)}^{ - 1}}{\pi ^{i,i}}\Omega _t^i\left( \ell  \right){\rm{\hat x}}_t^i\left( \ell  \right)} \right] + C  \label{eq:ORIC}
	\end{align}
	
	According to the \emph{Matrix Inversion Lemma}, we have
	
	\begin{align}
	{\left( {\Phi _t^i\left( \ell  \right)} \right)^{ - 1}} &= 2{\left[ {\Psi _t^i\left( \ell  \right){E^i}{{\left( {\Omega _t^i\left( {\ell  + 1} \right)} \right)}^{ - 1}}{{\left( {{E^i}} \right)}^T}\Psi _t^i\left( \ell  \right) - \Psi _t^i\left( \ell  \right)} \right]^{ - 1}}  \nonumber  \\
	& =  - 2\left\{ {{{\left( {\Psi _t^i\left( \ell  \right)} \right)}^{ - 1}} + {E^i}\left[ {\Omega _t^i\left( {\ell  + 1} \right) - {{\left( {{E^i}} \right)}^T}\Psi _t^i\left( \ell  \right){E^i}} \right]{{\left( {{E^i}} \right)}^T}} \right\}  \nonumber  \\
	& =  - 2\left\{ {{{\left( {\Psi _t^i\left( \ell  \right)} \right)}^{ - 1}} + {E^i}{{\left( {{\pi ^{i,i}}\Omega _t^i\left( \ell  \right)} \right)}^{ - 1}}{{\left( {{E^i}} \right)}^T}} \right\}
	\end{align} 
	
	And
	
	\begin{align}
	& {\left( {\Phi _t^i\left( \ell  \right)} \right)^{ - 1}}\Psi _t^i\left( \ell  \right){E^i}{\left( {\Omega _t^i\left( {\ell  + 1} \right)} \right)^{ - 1}}{\pi ^{i,i}}\Omega _t^i\left( \ell  \right){\rm{\hat x}}_t^i\left( \ell  \right)   \nonumber  \\
	& =  - 2\left\{ {{{\left( {\Psi _t^i\left( \ell  \right)} \right)}^{ - 1}} + {E^i}{{\left( {{\pi ^{i,i}}\Omega _t^i\left( \ell  \right)} \right)}^{ - 1}}{{\left( {{E^i}} \right)}^T}} \right\}\Psi _t^i\left( \ell  \right){E^i}{\left( {\Omega _t^i\left( {\ell  + 1} \right)} \right)^{ - 1}}{\pi ^{i,i}}\Omega _t^i\left( \ell  \right){\rm{\hat x}}_t^i\left( \ell  \right) \nonumber  \\
	& =  - 2\left\{ {{E^i}{{\left( {\Omega _t^i\left( {\ell  + 1} \right)} \right)}^{ - 1}}{\pi ^{i,i}}\Omega _t^i\left( \ell  \right){\rm{\hat x}}_t^i\left( \ell  \right)} \right. \nonumber  \\
	& \;\;\;\; + \left. {{E^i}{{\left( {{\pi ^{i,i}}\Omega _t^i\left( \ell  \right)} \right)}^{ - 1}}\left[ {\left( {\Omega _t^i\left( {\ell  + 1} \right)} \right) - {\pi ^{i,i}}\Omega _t^i\left( \ell  \right)} \right]{{\left( {\Omega _t^i\left( {\ell  + 1} \right)} \right)}^{ - 1}}{\pi ^{i,i}}\Omega _t^i\left( \ell  \right){\rm{\hat x}}_t^i\left( \ell  \right)} \right\} \nonumber  \\
	& =  - 2{E^i}{\rm{\hat x}}_t^i\left( \ell  \right)
	\end{align}
	
	Consequently, (\ref{eq:ORIC}) can be rewritten as
	
	\begin{align}
	J_t^i\left( {{\Theta ^i},\ell } \right) &= {\left[ {{\Theta ^i} + {\bf{x}}_t^i\left( \ell  \right) - {E^i}{\rm{\hat x}}_t^i\left( \ell  \right)} \right]^T}\Phi _t^i\left( \ell  \right)\left[ {{\Theta ^i} + {\bf{x}}_t^i\left( \ell  \right) - {E^i}{\rm{\hat x}}_t^i\left( \ell  \right)} \right]  \nonumber  \\
	& =  (\Theta^i)^T \Phi^i_t(\ell)  \Theta^i + 2 (\Theta^i)^T  \varphi^i_t(\ell) + c^i_t(\ell)
	\end{align}
	
	As the result, (\ref{OmeA}) and (\ref{PsiA}) are proved.
	
\end{proof}

In order to obtain a reliable estimate of the drift parameters, we follow the strategy usually adopted in recursive parameter estimation and consider a total loss function up to the current time instant/consensus step defined recursively as

\[
\bar J^i_t(\Theta^i , \ell+1) = \lambda \bar J^i_t(\Theta^i , \ell) + J^i_t(\Theta^i , \ell)
\]
where, in accordance with the algorithm of Table I, we have $\bar J^i_{t+1}(\Theta^i , 0) = \bar J^i_t(\Theta^i , L) $ and the scalar $\lambda \in (0,1)$ is a forgetting factor used to discount the past (since the quality of the local estimate of the state vector improves with time,
it is reasonable to weight less the past loss functions). As it can be easily checked, recursive minimization of $\bar J^i_t(\Theta^i , \ell)$ gives rise to a standard recursive least squares (RLS) algorithm of the form
\begin{eqnarray}
\hat \Theta^i_t (\ell +1)  &=& \hat \Theta^i_t (\ell )  - [\bar  \Phi^i_t(\ell+1)]^{-1} [ \varphi^i_t(\ell) + \Phi^i_t(\ell) \hat \Theta^i_t (\ell )  ] \nonumber \\
\bar  \Phi^i_t(\ell+1) &=& \lambda \bar  \Phi^i_t(\ell) + \Phi^i_t(\ell) \, . \label{eq:RLS}
\end{eqnarray}

Combining the recursive drift parameter estimation algorithm (\ref{eq:RLS}) with the consensus Kalman filter of Table I, we get the JTTSL-CSKF
algorithm reported in Table \ref{JCKFCI}.

\begin{remark}
	When the computational capabilities of each single sensor are low, several simplifications of the proposed solution are possible.
	For instance, instead of using an RLS algorithm for updating the estimate which requires the storage of a possibly large matrix,
	we can simply use an online optimization technique of the form
	
	\begin{equation}   \label{ODOC}
	\hat \Theta^i_t (\ell +1) = \hat \Theta^i_t (\ell ) - \gamma_t(\ell) \,\left .  \nabla_{\Theta^i}  J^i_t(\Theta^i , \ell) \right |_{\Theta^i  = \hat \Theta^i_t (\ell ) }
	\end{equation}
	where $\gamma_t(\ell) $ is a suitable stepsize.
	Another possible simplification in order to alleviate the computational burden amounts to updating the estimates of the drift parameters only once at each sampling interval, for instance ofter the last consensus step $L$.
	This approximated solution has been employed in the simulation results of the following section.
	
\end{remark}

\begin{table}
	\caption{JTTSL-CSKF ( node $i$, time $t$ )}   \label{JCKFCI}
	\begin{center}
		\begin{tabular}{l p{5cm}|}
			\hline \hline
			\underline{Correction} \\ \hline
			$\;\;$ Perform the correction step as in table I \\  \hline
			\hline
			\underline{Consensus and calibration} \\ \hline
			$\;\;$ Set  $\hat \Theta^i_{t}(0) = \hat \Theta^i_{t-1}(L) $ \\
			$\;\;$ Set  $\bar \Phi^i_{t}(0) = \bar \Phi^i_{t-1}(L) $ \\
			$\;\;$ For $\ell = 0,1, \ldots ,L - 1$ do \\
			$\;\;\;\;$  Exchange the information with its neighbors and  \\
			$\;\;\;\;$  compute $\hat \Theta^i_{t}(\ell+1)$  and  $\bar \Phi^i_{t}(\ell+1)$ as in (\ref{eq:RLS}) \\
			$\;\;\;\;$  Fuse the quantities $q_t^i(\ell)$ and $\Omega _t^i(\ell)$ according to \\
			$\;\;\;\;\;\;$ $q_t^i(\ell + 1) = \sum\limits_{j \in {{\cal N}^i}} {{\pi ^{i,j}}[q_{\left. t \right|t}^j(\ell) + \Omega _{\left. t \right|t}^j(\ell) \, {\hat \theta^{i,j}_t (\ell+1)}]}$  \\
			$\;\;\;\;\;\;$ $\Omega _t^i(\ell + 1) = \sum\limits_{j \in {{\cal N}^i}} {{\pi ^{i,j}}\Omega _{\left. t \right|t}^j(\ell)}$  \\
			$\;\;$ End for \\
			$\;\;$ $\hat {\rm x}_{\left. t \right|t}^i = {[\Omega _{\left. t \right|t}^i(L)]^{ - 1}}q_t^i(L),\;\;\Omega _{\left. t \right|t}^i = \Omega _t^i(L)$  \\ \hline
			\underline{Prediction}  \\ \hline
			$\;\;$ Perform the prediction step as in table I \\
			\hline \hline
		\end{tabular}
	\end{center}
\end{table}

\section{Simulation examples}

In this section, the performance of the proposed JTTSL-CSKF algorithm is evaluated by means of simulation experiments.
The target motion obeys the constant velocity model, which means that
${f_t}\left( {{{\mathop{\rm x}\nolimits} _t^i}} \right) = A {{\mathop{\rm x}\nolimits} _t^i}$ in (\ref{ds}) and matrices $A, Q$ are given by
\begin{equation}   \label{TDM}
A = \left( {\begin{array}{*{20}{c}}
	1&T&0&0\\
	0&1&0&0\\
	0&0&1&T\\
	0&0&0&1
	\end{array}} \right),  \,Q = \sigma _x^2 I_4
\end{equation}
where ${\sigma _x} = 10 \, [m]$, $T=1 \, [s]$, and $\rm{x}_t^i$ represents the target state with respect to a coordinate system with origin at the location of node $i$.
The algorithm is tested on both linear and nonlinear sensors with measurement functions in (\ref{mf})  given as follows

\underline{Linear sensor}: $h_t^i\left( {{\mathop{\rm x}\nolimits} _t^i} \right) = C^i {\mathop{\rm x}\nolimits} _t^i$ where

\begin{equation}   \label{LOM}
C^i = {\alpha ^i}\left( {\begin{array}{*{20}{c}}
	1&0&0&0\\
	0&0&1&0
	\end{array}} \right), R^i = \sigma _y^2 I_2
\end{equation}

\begin{figure}[h]
	\centering {
		\begin{tabular}{ccc}
			\includegraphics[width=1\textwidth]{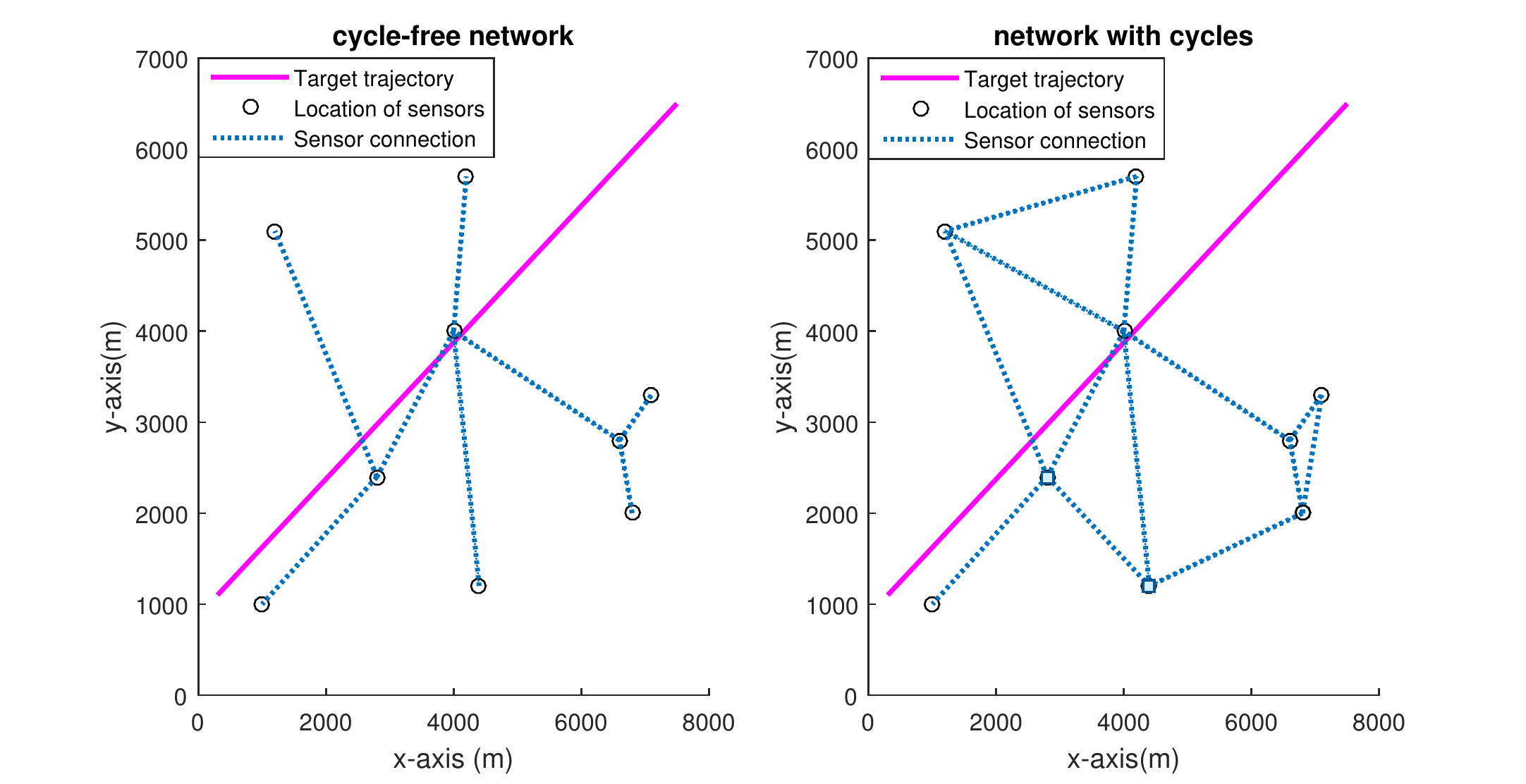}\\
		\end{tabular}
	}
	\caption{Tree network and network with cycles.}
	\vspace{-0.5\baselineskip}
	\label{topology}
\end{figure}

\underline{Nonlinear sensor}:

\begin{equation}   \label{NLOM}
h_t^i\left( {{\mathop{\rm x}\nolimits} _t^i} \right){\rm{ = }}\left( {\begin{array}{*{20}{c}}
	{\sqrt {{{\left( {\xi_t^i} \right)}^2} + {{\left( {\eta_t^i} \right)}^2}} }\\
	{{\mathop{\rm atan}\nolimits} ( {{\eta_t^i} \mathord{\left/
				{\vphantom {{\eta_t^i} {\xi_t^i}}} \right.
				\kern-\nulldelimiterspace} {\xi_t^i}} ) }
	\end{array}} \right), \, R^i = {\mathop{\rm diag}\nolimits}(\sigma _r^2,\sigma _\beta ^2)
\end{equation}
where ${\alpha ^i}$ are randomly selected for each node within the interval $[0.75,1.25]$, ${\sigma _y} = 30 \,[m],
{\sigma _r} = 15 \, [m]$, and ${\sigma _\beta } = {0.5 \,[^o]}$.
The proposed algorithm is also tested on two different types of networks: one with  a tree topology and the other with cycles, both
consisting of $9$ nodes as shown in Fig. \ref{topology}.
The initial drift parameter estimates are set to $\hat \theta _0^{i,j} = 0,\left( {i,j} \right) \in {\cal A}$, in both networks.
The stepsize is set as ${\gamma _t} = 1.2$ for the linear sensor case and, respectively, ${\gamma _t} = 20$ for the nonlinear sensor case.

\begin{figure}[h]
	\centering {
		\begin{tabular}{ccc}
			\includegraphics[width=1\textwidth]{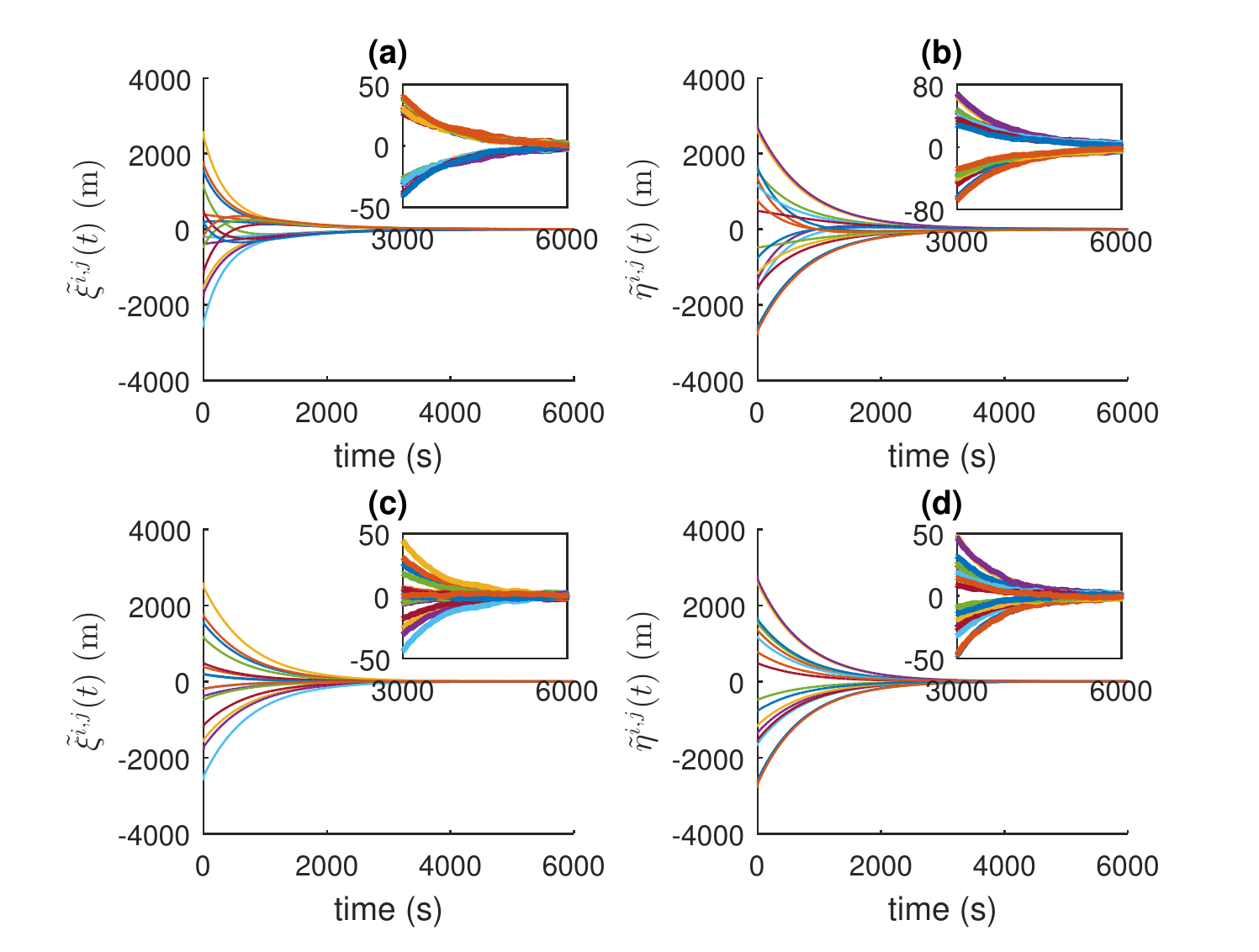}\\
		\end{tabular}
	}
	\caption{Time-behavior of ${{\tilde \theta }^{i,j}}(t),\left( {i,j} \right) \in {\cal A}$, in a tree network with linear sensors, for $L=1$ (a), (b) and $L=10$  (c), (d) consensus steps.}
	\vspace{-0.5\baselineskip}
	\label{thetadiflineartree}
\end{figure}

\begin{figure}[h]
	\centering {
		\begin{tabular}{ccc}
			\includegraphics[width=1\textwidth]{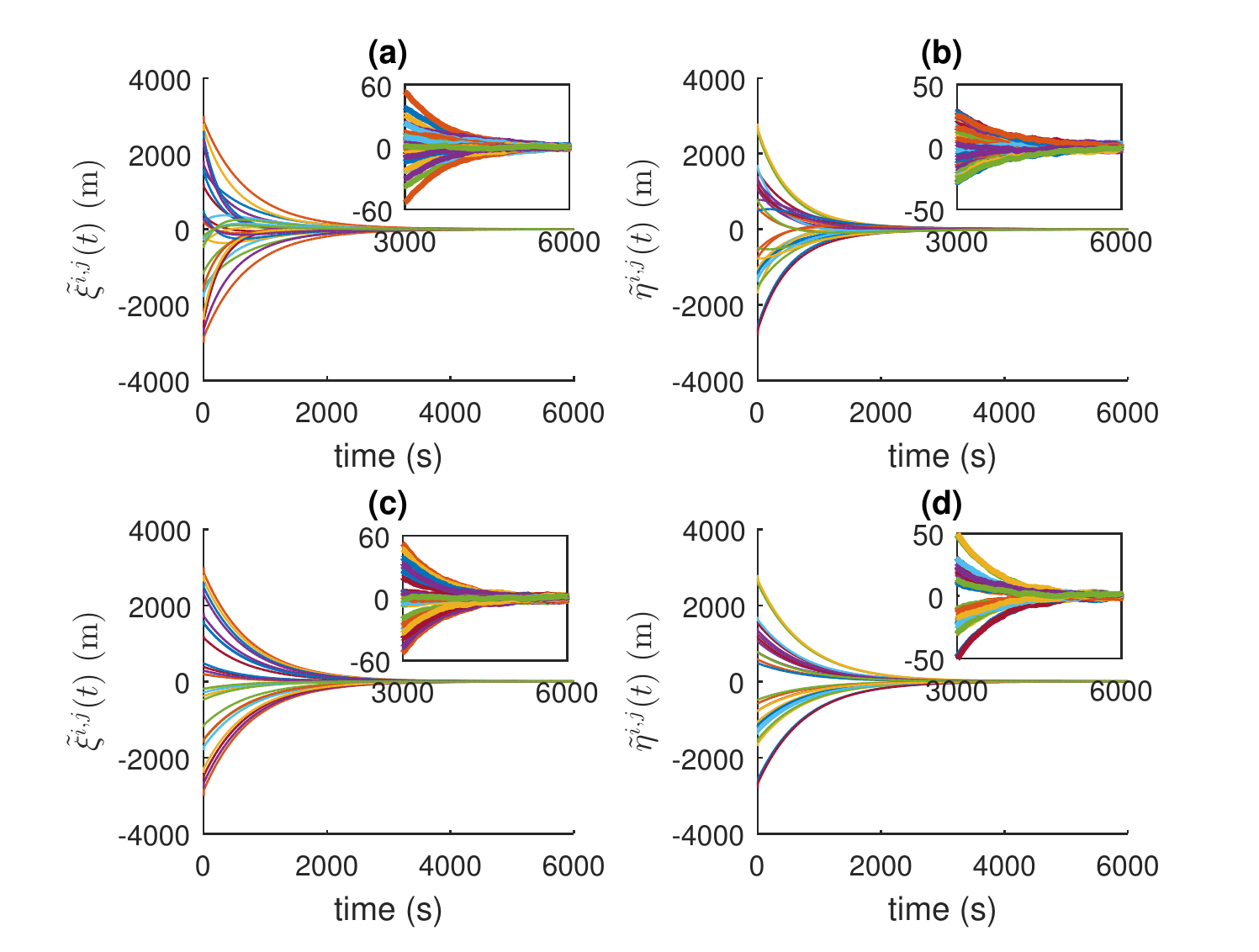}\\
		\end{tabular}
	}
	\caption{Time-behavior of ${{\tilde \theta }^{i,j}}(t),\left( {i,j} \right) \in {\cal A}$, in a network with cycles and linear sensors, for $L=1$ (a), (b) and
		$L=10$ (c), (d) consensus steps.}
	\vspace{-0.5\baselineskip}
	\label{thetadiflinearcircle}
\end{figure}

\begin{figure}[h]
	\centering {
		\begin{tabular}{ccc}
			\includegraphics[width=1\textwidth]{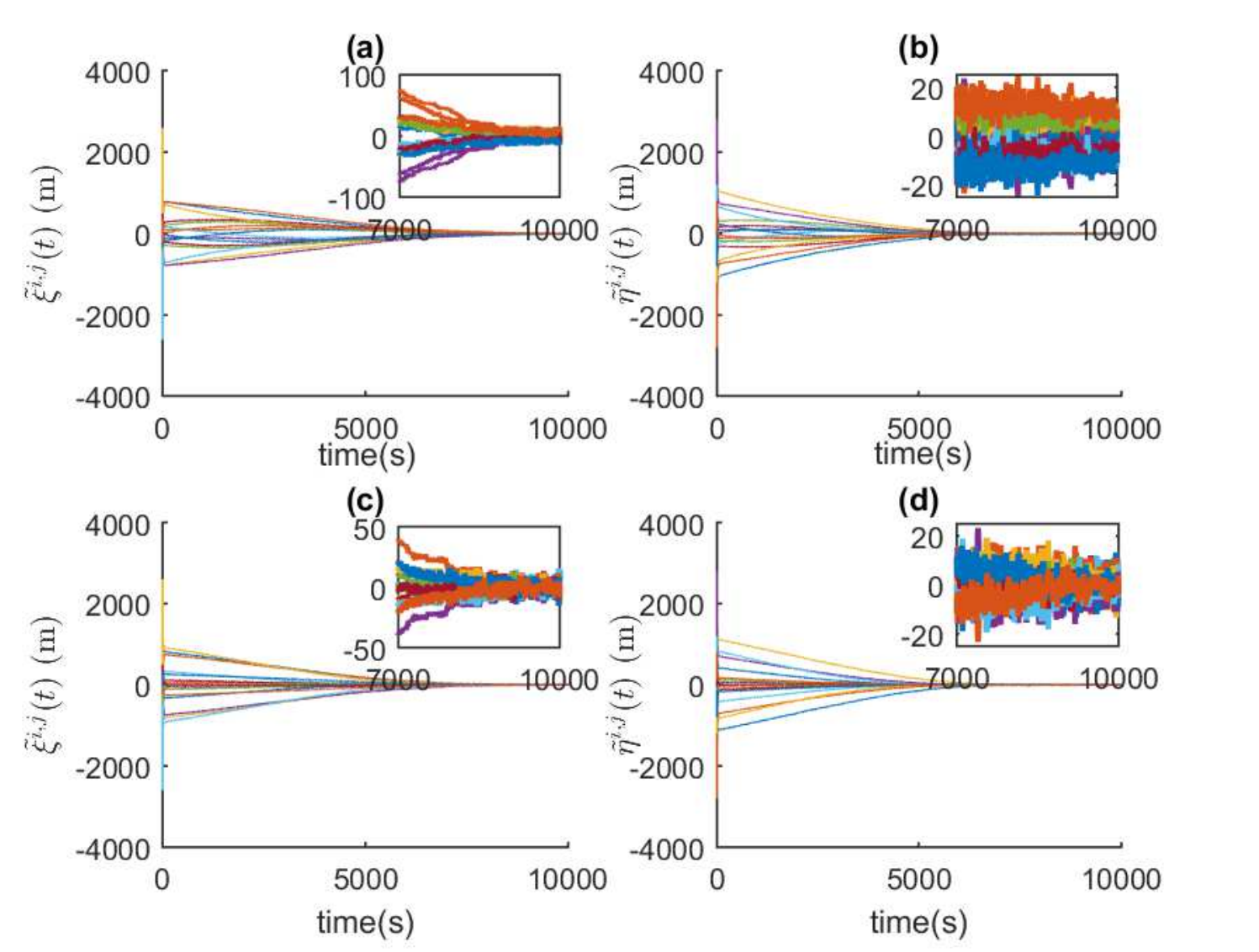}\\
		\end{tabular}
	}
	\caption{Time-behaviour of ${{\tilde \theta }^{i,j}}(t),\left( {i,j} \right) \in {\cal A}$, in a tree network with nonlinear sensors, for $L=1$ (a), (b) and $L=10$
		(c), (d) consensus steps.}
	\vspace{-0.5\baselineskip}
	\label{thetadifnonlineartree}
\end{figure}

\begin{figure}[h]
	\centering {
		\begin{tabular}{ccc}
			\includegraphics[width=1\textwidth]{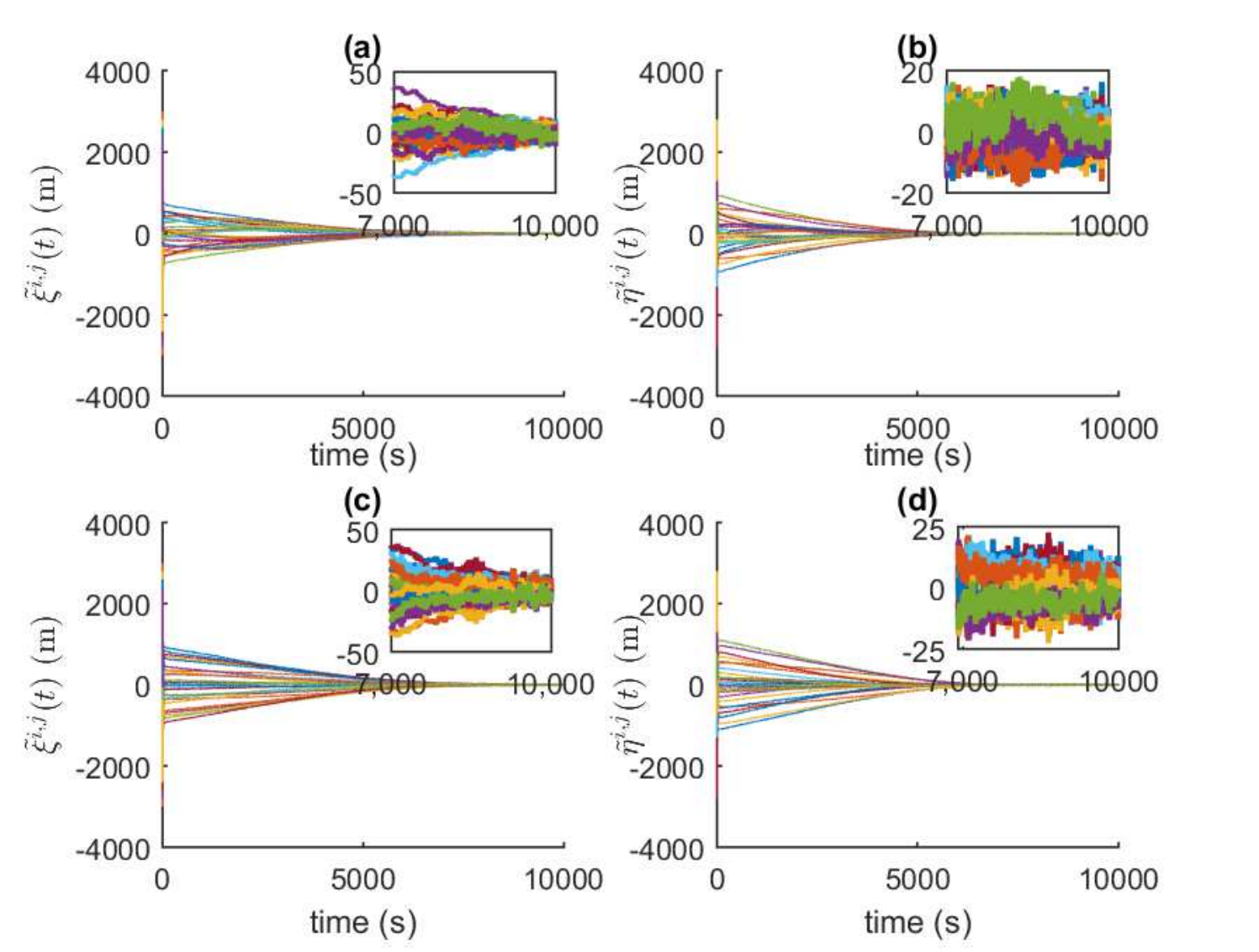}\\
		\end{tabular}
	}
	\caption{Time-behavior of ${{\tilde \theta }^{i,j}}(t),\left( {i,j} \right) \in {\cal A}$, in a network with cycles and nonlinear sensors, for $L=1$ (a), (b) and $L=10$ (c), (d) consensus steps.}
	\vspace{-0.5\baselineskip}
	\label{thetadifnonlinearcircle}
\end{figure}

\begin{figure}[h]
	\centering {
		\begin{tabular}{ccc}
			\includegraphics[width=1\textwidth]{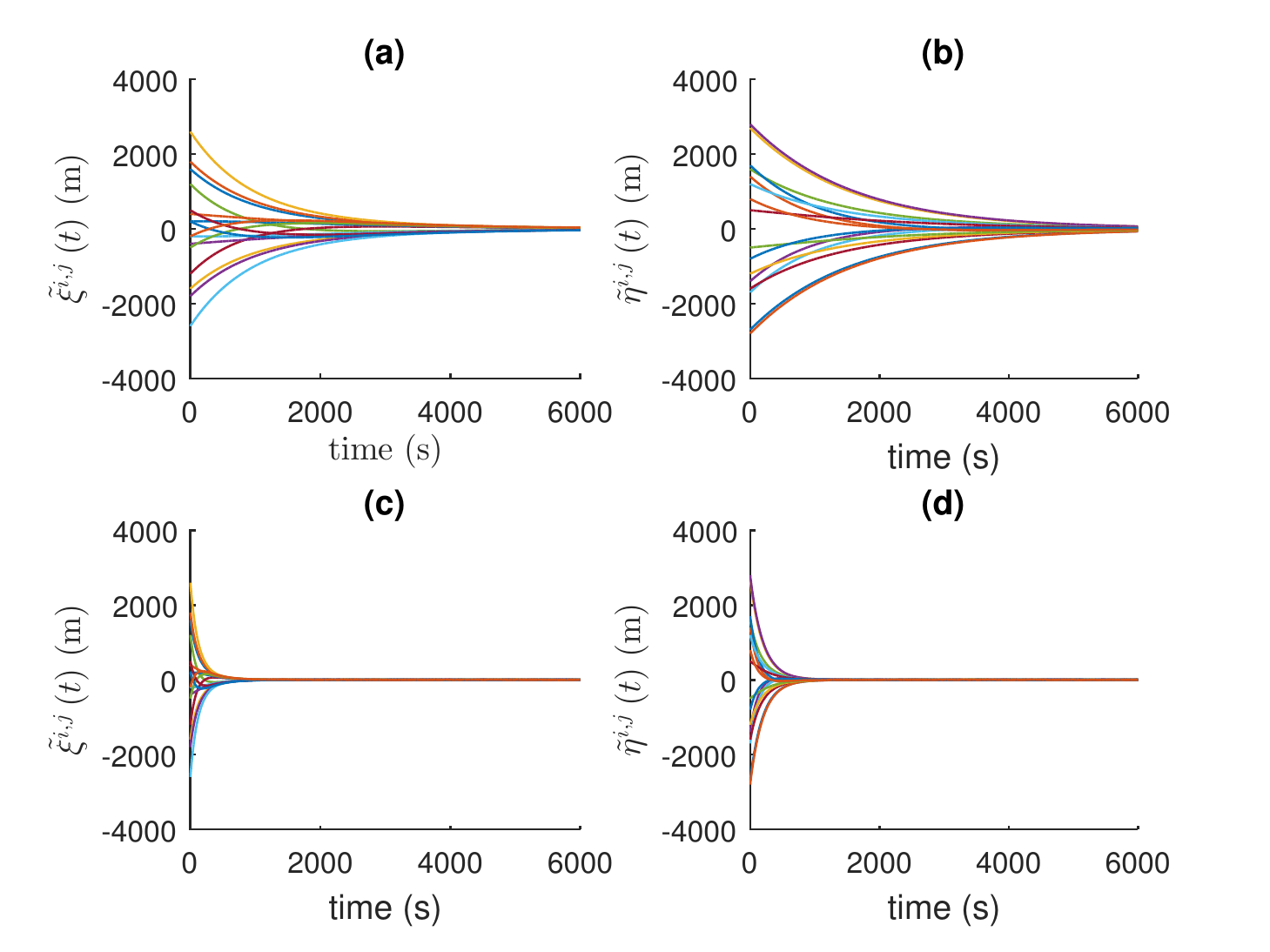}\\
		\end{tabular}
	}
	\caption{Stepsize affections of the speed of convergence, for $\gamma_t = 0.6$ (a),(b) and $\gamma_t = 4.8$ (c),(d).}
	\vspace{-0.5\baselineskip}
	\label{diffstepsize}
\end{figure}

Figs. \ref{thetadiflineartree}-\ref{thetadifnonlinearcircle} analyse the self-localization performance by displaying the time evolution, in single typical realizations, of the drift parameter estimation errors
$\tilde{\theta}^{i,j}_t  \buildrel \Delta \over  = \theta^{i,j} - \hat{\theta}^{i,j}_t  = \left[ \tilde{\xi}_t^{i,j}, 0, \tilde{\eta}^{i,j}_t, 0 \right]^T$ for the two considered networks of Fig.
\ref{topology} and the two types of sensors in (\ref{LOM}) and (\ref{NLOM}).
It can be seen that in all cases the estimated drift parameters converge to their true values.
Further, the effect of the stepsize $\gamma_t$ on the speed of convergence is shown in Fig. \ref{diffstepsize} in the case of tree network and linear sensors for $L=1$. 
It can be seen that the speed of self-localization of JTTSL-CSKF can be controlled by tuning $\gamma_t$, thus providing robustness in real applications.


\begin{figure}[h]
	\centering {
		\begin{tabular}{ccc}
			\includegraphics[width=1\textwidth]{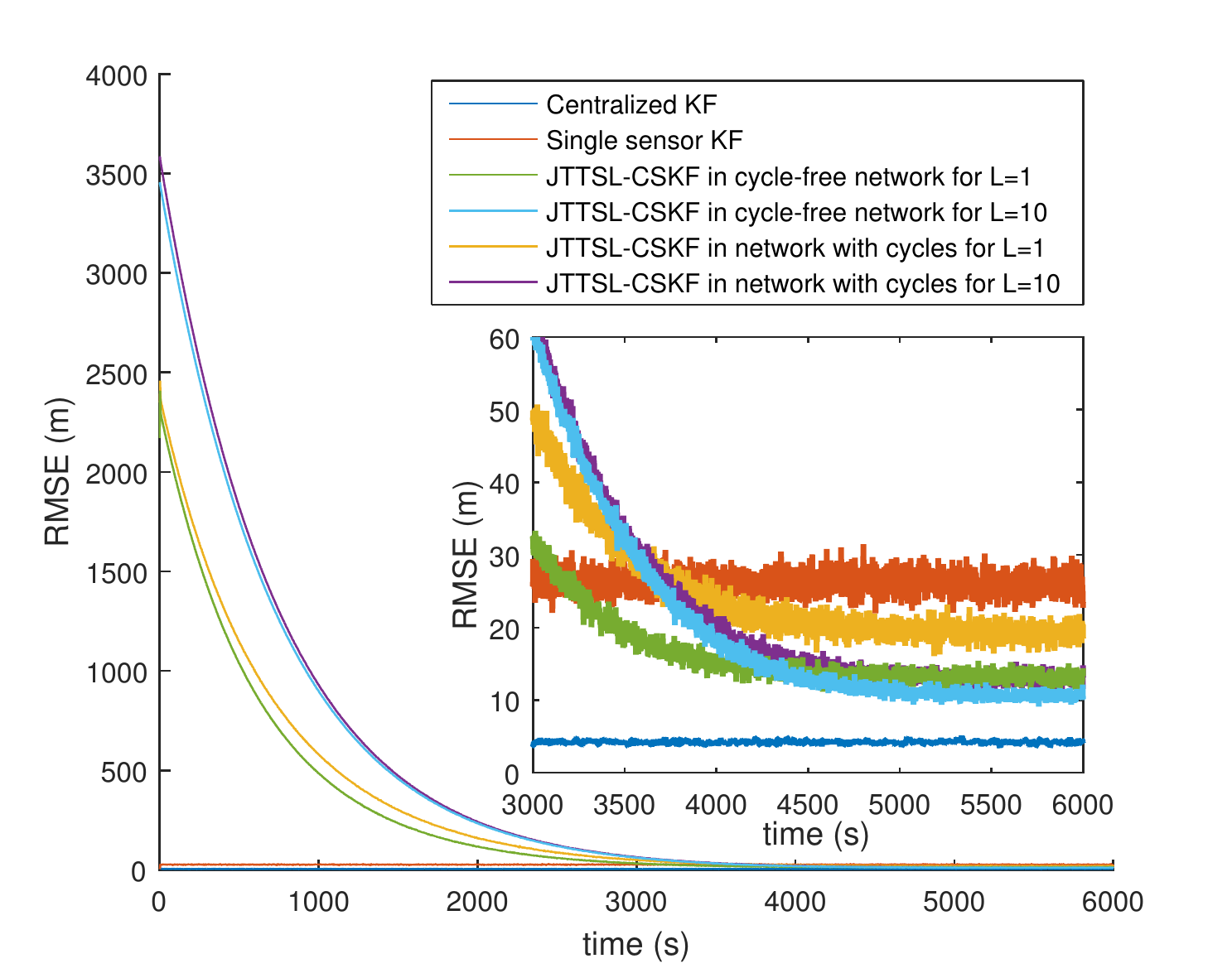}\\
		\end{tabular}
	}
	\caption{Time-behavior of the target state RMSE under linear observations.}
	\vspace{-0.5\baselineskip}
	\label{RMSElinear}
\end{figure}

\begin{figure}[h]
	\centering {
		\begin{tabular}{ccc}
			\includegraphics[width=1\textwidth]{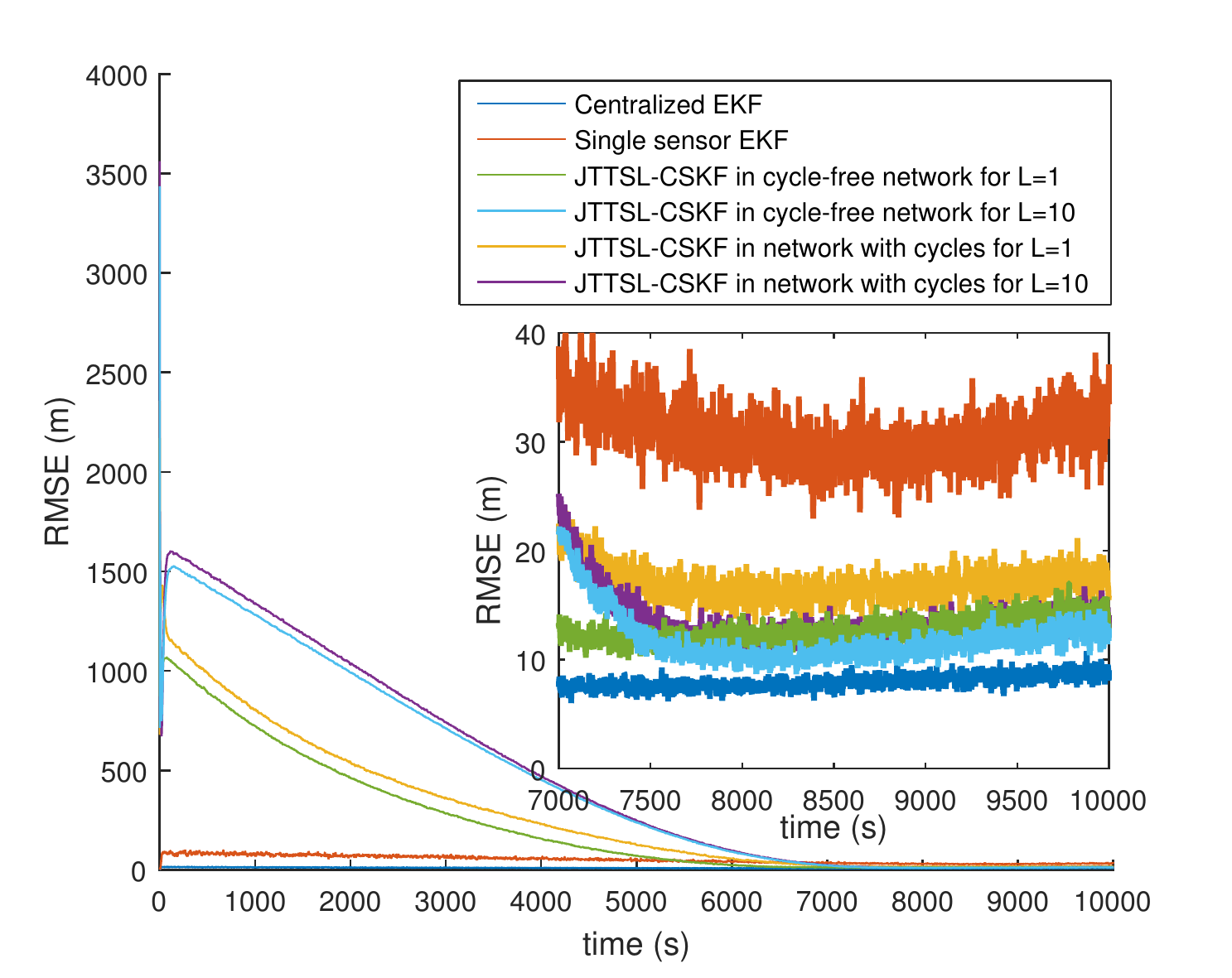}\\
		\end{tabular}
	}
	\caption{Time-behavior of the target state RMSE under nonlinear observations.}
	\vspace{-0.5\baselineskip}
	\label{RMSEnonlinear}
\end{figure}

Next, Figs. \ref{RMSElinear} and \ref{RMSEnonlinear} analyse the target tracking performance by showing  the time evolution of the
target state Root Mean Square Error (RMSE), averaged over $200$ independent Monte Carlo trials,
in the two cases of linear (\ref{LOM}) and, respectively, nonlinear sensors (\ref{NLOM}).
In both figures, the RMSEs of the centralized (Extended) KF and the single-sensor (Extended) KF are also reported as reference lower and upper bounds for the performance of 
the JTTSL-CSKF algorithm.
Further, performance of JTTSL-CSKF is also reported for the two network scenarios  of Fig. \ref{topology} as well as for
a single ($L=1$) and multiple ($L=10$) consensus steps.
It can be seen that, at convergence, the performance of the proposed method stays in between the centralized and single sensor (E)KF for any $L$.
Further, as expected, the RMSE of the proposed method is getting closer to the centralized (E)KF when the number $L$ of consensus steps is increased. 
It can also be seen that the proposed method has wide potential applications, being immune to the type of network.

A final consideration is important. By looking at the simulation results, it is evident that during the initial learning phase
the JTTSL-CSKF performs worse than the single-sensor filter. This is due to the fact that, at the beginning, the estimates
of the drift parameters are not accurate enough and hence the fusion is not reliable.
This drawback can be easily avoided with a slight  modification of the proposed approach in which the fusion is performed at each time instant in order to update
the estimates of the drift parameters, but the local estimates of the target state are replaced with the fused ones only when the drift parameter estimates are reliable enough 
(for example, when the loss function $J^i_t(\Theta^i,\ell)$ is below a given threshold).


\section{Conclusion}
The paper has addressed the important practical issue of fusing, in a distributed way,  target tracking information over a sensor network with sensor nodes that get target measurements in their own local coordinate system and do not know their relative positions.
The problem, referred to as \textit{Joint Target Tracking and Sensor self-Localization} (JTTSL), has been solved by exploiting a novel computationally efficient
consensus Kalman filtering approach by which each node is able to self-localize with respect to neighbors and thus correctly fuse target-state information.
The effectiveness of the proposed approach has been successfully tested via simulation experiments concerning both networks with tree topology and networks with
cycles as well as both linear and nonlinear sensors.
Future developments of this work will possibly regard a theoretical convergence analysis as well as extensions in several directions like, e.g.,
multitarget scenario, mobile sensors, the case of unknown relative orientations among sensors and cooperative multirobot simultaneous localization and mapping.

\label{--------------------------------------------sec: End---------------------------------------------------}

\end{document}